\documentclass[fleqn,10pt]{wlscirep}
\usepackage[utf8]{inputenc}
\usepackage[T1]{fontenc}
\usepackage{marvosym}
\usepackage{amsmath}
\usepackage{nccmath}

\title{Generation of arbitrary abruptly autofusing circular Airy Gaussian vortex vector modes
}

\author[1,*]{Xiao-Bo Hu}
\author[2]{Bo-Zhao}
\author[1]{Rui-Pin Chen}
\author[3,*]{Carmelo Rosales-Guzm\'an}
\affil[1]{Key Laboratory of Optical Field Manipulation of Zhejiang Province,Department of Physics, Zhejiang Sci-Tech University, Hangzhou, 310018, China}
\affil[2]{Wang Da-Heng Collaborative Innovation Center for Quantum manipulation \& Control, Harbin University of Science and Technology, Harbin 150080, China}
\affil[3]{Centro de Investigaciones en Óptica, A.C., Loma del Bosque 115, Colonia Lomas del campestre, C.P. 37150 León, Guanajuato, Mexico}

\affil[*]{huxiaobo@zstu.edu.cn}
\affil[*]{carmelorosalesg@cio.mx}

\begin{abstract}
Complex vector modes represent a general state of light nonseparable in their spatial and polarization degrees of freedom, which have inspired a wide variety of novel applications and phenomena, such as their unexpected propagation behaviour. For example, they can propagate describing periodic polarization transitions, changing from one vector beam to another. Here, we put forward a novel class of vector modes with the capability to experience an abruptly autofocusing behaviour. To achieve such beams, we encode the spatial degree of freedom in the Circular Airy Gaussian vortex (CAGV) beams. We demonstrate the experimental generation of arbitrary CAGV vector beams and evince some of their properties, such as a rotation of intermodal phase. We anticipate that the fascinating properties of theses modes will prompt the development of novel applications associated to their autofocusing behaviour and polarization distribution. 
\end{abstract}
\begin{document}

\flushbottom
\maketitle
%
%
\section*{Introduction}
Complex vector light beams have drawn a significant amount of attentions in recent time, in part due to the many applications that have inspired, but also due to the novel concepts and phenomena that are giving rise to \cite{Rosales2018Review,Roadmap}. Such modes are generated as a non-separable superposition between the spatial and polarization degrees of freedom (DoF), in a mathematical form identical to that of entangled photons, from which they coined the term classically-entangled. Crucially, the unlimited freedom in the spatial DoF gives rise not only to exotic polarization and intensity distributions in the transverse plane, but also to intriguing propagation behaviours, such as periodic or non-periodic polarization transitions, self-healing, transverse acceleration or resilience against atmospheric turbulence  \cite{Otte2018,Zhong2021,Li2017,Otte2018Recovery,Nape2018,Nape2022,Hu2021,Zhaobo2022}. Even though cylindrical vector modes, such as, Laguerre- or Bessel-Gaussian, have gained popularity over the years, in recent time the generation of vector beams in noncylindrical coordinates systems have become a topic of late \cite{Hu2021,Zhaobo2022,Rosales2022,YaoLi2020}. For example, travelling Parabolic- or accelerating-Gaussian vector modes that upon propagation in free space, evolve from a pure vector state to a quasi scalar one \cite{Hu2021} or accelerate along a parabolic trajectory preserving their polarization structure \cite{Zhaobo2022}. Vector beams have already demonstrated their potential in a wide variety of research areas such as, optical tweezers, optical metrology, optical communications, among others \cite{Toppel2014,Yuanjietweezers2021,Hu2019,Milione2015,Rosales2018Review,Shen2022,Ndagano2018,Ndagano2017,Fang2021}. Hence, along with the discovery of novel vector beams, other fascinating properties will be discovered, paving the path to other applications.\\
As such, in this manuscript, we propose and experimentally demonstrate a general  class of vector modes, which we term  Circular Airy Gaussian vortex vector (CAGVV) beams, that are generated from a non-separable superposition of the polarization and spatial DoF encoded in the set of orthogonal Circular Airy Gaussian Vortex (CAGV) beams. Even though CAGV modes are well-known due to their abrupt autofocusing dynamics along the propagation direction, which has encountered several applications \cite{Efremidis2019}, most of the work has been carried out in the scalar regime, including Circular Airy Gaussian beams, CAGV, elliptical Airy beams, amongst others \cite{Papazoglou:11,Jiang2012,Zha:18}. Despite that a similar class of vector beams has been reported in the past \cite{Liu2013}, the authors only focused on the abrupt polarization transition. Here we not only propose the generation of arbitrary CAGVV modes but also analyze the phase dynamics that such modes of different topological charges undergo upon propagation. More precisely, their intermodal phase rotates upon propagation as a function of the topological charges constituiting the CAGVV vector mode. We provide a mathematical expression for this intermodal phase shift and demonstrate it experimentally. Finally, a propagation-dependent transition from linear to elliptical polarization is evinced through the Stoke Parameter $S_3$, which we corroborate theoretically and experimentally. Due to their intriguing properties, we anticipate CAGVV beams to be a practical and flexible tool, with applications in both, fundamental and applied research.
\section*{THEORETICAL BACKGROUND}\label{THEORETICAL BACKGROUND}

In the cylindrical polar coordinates, CAGV beams are given by, 
\begin{ceqn}
\begin{equation}
    \centering
     {\bf U}_{m}({\bf r},\phi)={r^{\lvert m \rvert}}Ai\left(\frac{r_0-r}{\omega} \right)\textrm{e}^{ a(r_0-r)/\omega}\textrm{e}^{im\phi}\textrm{e}^{i\nu r},
    \label{Eq:CAV}
\end{equation}
\end{ceqn}
where $Ai({\cdot})$ is the Airy function, $\omega$ is a scaling factor, $a$ is a truncation parameter and $r_0$ is the beam radius in the initial plane. The parameter $\nu$ is the initial launch angle, which plays a key role in the parabolic trajectory the beam follows upon propagating, for $\nu>0$ propagate outwards along a diverging parabolic trajectory, while for $\nu<0$ they propagate inwards describing an abruptly autofocusing trajectory. Additionally, such CAGV beam features an azimuthally-varying phase of the form $\exp({im\phi})$, where the index $m \in \mathbb{Z}$ is known as the topological charge, associated to an $m\hbar$ amount of orbital angular momentum(OAM) per photon, with $\hbar$ being the reduced Plank constant.

To generate the CAGVV modes we coaxially superimpose two orthogonal CAGV modes of different topological charges and orthogonal polarization state in a nonseparable weighted superposition\cite{Galvez2012,Galvez2015}. Such superposition can be mathematically  written as,

\begin{ceqn}
\begin{equation}
    \centering
    \Psi_{m_1, m_2}({\bf r},\phi)=\cos\theta {\bf U}_{m_1}({\bf r},\phi)\hat{\bf e}_R
  +\sin\theta\exp(i\alpha) {\bf U}_{m_2}({\bf r},\phi)\hat{\bf e}_L,
    \label{Eq:VM}
\end{equation}
\end{ceqn}
where $\hat{\bf e}_R$ and $\hat{\bf e}_L$ are unitary vectors representing the right- and left-handed circular polarization states, respectively. ${\bf U}_{m_1}({\bf r},\phi)$ and ${\bf U}_{m_2}({\bf r},\phi)$ represent the CAGV modes with topological charges $m_1$ and $m_2$, respectively, weighted by the parameter $\theta\in[0,\pi/2]$. Additionally, the inter-modal phase $\exp(i\alpha)$, with $\alpha\in [0, \pi]$, sets a phase delay between both polarization components. From now on, we will omit the explicit dependence of $({\bf r},\phi)$, unless is necessary.

\begin{figure}[tb]
    \centering
    \includegraphics[width=0.880\textwidth]{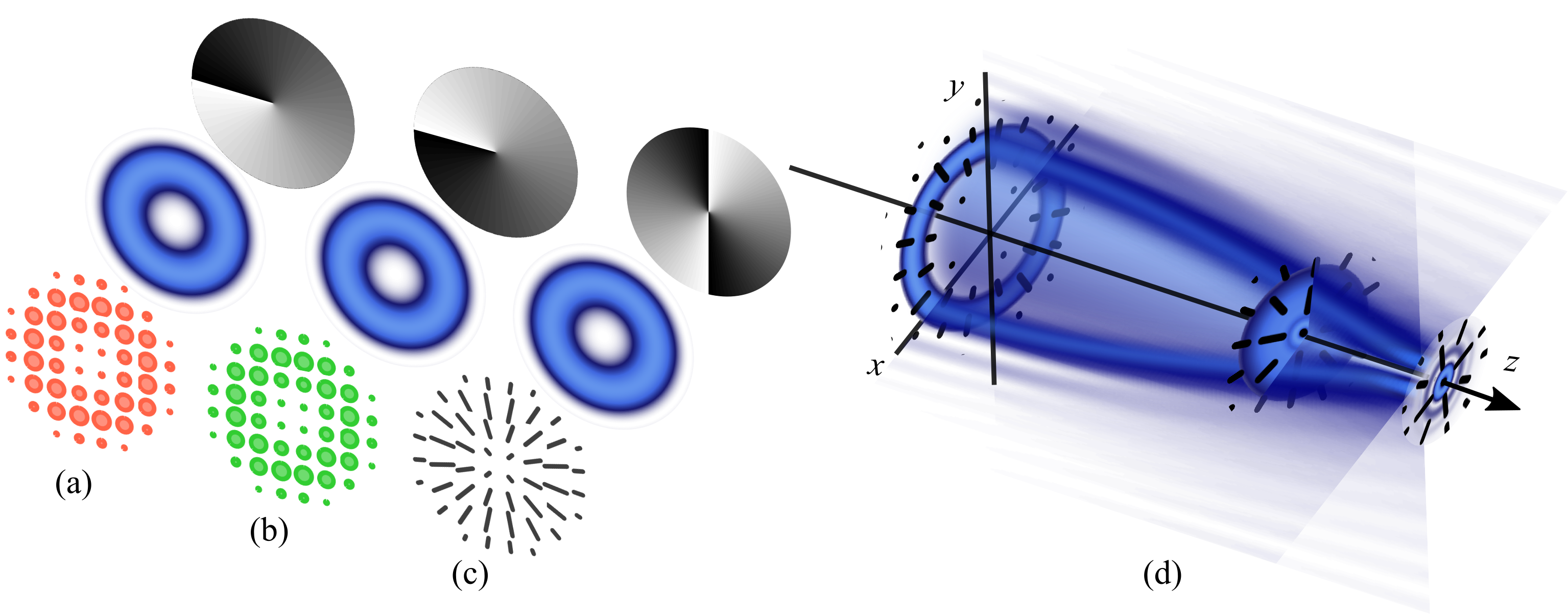}
    \caption{Polarization distribution (front panels), intensity profiles (middle panels) and transverse phase (back panels) of the modes (a)${\bf U}_{1}\hat{\bf e}_R$, (b) ${\bf U}_{-1}\hat{\bf e}_L$ and (c) $\Psi_{1,-1}$, respectively. In (d) we illustrate schematically the autofocusing effect such beams experience upon propagation. Here, orange and green ellipses represent right and left circular polarization, while black lines symbolize linear polarization. For this example, a=1, w=0.1, $r_0$=1, $\nu_1=\nu_2=0$.}
    \label{concept}
\end{figure}

For the sake of clarity, such superposition is schematically shown in Fig. \ref{concept} for the specific case $\Psi_{1,-1}$, constituted by the superposition of the scalar modes ${\bf U}_{1}\hat{\bf e}_R$ and ${\bf U}_{-1}\hat{\bf e}_L$, as shown in Fig.\ref{concept}(a) and Fig.\ref{concept}(b), respectively. Here, their polarization distribution is shown in the front panels with the orange ellipses representing right-handed circular polarization and green left circular. The middle panels show their intensity profile and the back panels their respective phase distribution. For this specific example we selected $\theta=\pi/4$, $\alpha=0$ and $z=0$, the resulting vector mode features a transverse radial polarization distribution, as schematically shown in Fig.\ref{concept}(c).

Importantly, the intensity profile of CAGVV beams inherit the self-focusing property of CAGV modes and therefore also feature an abruptly intensity collapse as function of propagation, as schematically illustrated in Fig.\ref{concept}(d). In addition their intermodal phase undertakes a phase rotation $ \Delta\phi$, which depends directly on the topological charge of both CAGV scalar components. Such phase rotation has been observed theoretically before and according to Dong Ye {\it el. al.}, it is not attributed to the Gouy phase optical vortices acquire upon propagation in free space \cite{Ye2017}. Crucially, it was demonstrated in the context of scalar modes that a coaxial superposition of Laguerre-Gaussian beams give rise to off-axis optical vortices, which upon propagation rotate about the beam axis due to the Gouy phase. Such rotation is proportional to the topological charges of the constituting beams, which for the specific case $p=0$ reduces to\cite{Baumann2009},  
\begin{ceqn}
\begin{equation}
 \centering
    \Delta\phi=\frac{|m_2|-|m_1|}{m_2-m_1}\Delta\xi,
    \label{Gou}
\end{equation} 
\end{ceqn}
where $\Delta\xi=\arctan(z/z_R)$ and $z_R$ is the Rayleigh distance. We propose and demonstrate experimentally that even though the constituting scalar beams in CAGVV modes do not interfere since they have orthogonal polarization, the rotation of the intermodal phase can also be described by Eq. \ref{Gou}.


\section*{Results}
We now conduct a detailed analysis of the propagation dynamics of the generated CAGVV beams, both experimentally and through numerical simulations. More specifically, we first compute the four Stokes parameters and from these reconstruct the transverse polarization distribution of CAGVV beams. From this we can also reconstruct the intermodal phase as $\alpha=\arctan(S_2/S_1)$ and monitor its rotation as function of propagation. Finally, through the Stokes parameter $S_3$ we can monitor the evolution of circular polarization as the mode propagates. Such analysis is carried out for three different cases, namely, $|m_1|=|m_2|$, $|m_1|>|m_2|$ and $|m_1|<|m_2|$, at the planes given by $z=0.0$ mm, $z=720.0$ mm, and $z=783.0$ mm. All beams were generated using the specific parameters $a=0.4$, $w=0.1$, $r_0$=1, $\nu_1=\nu_2=0$, and $\theta=\pi/4$. Applying Eq. \ref{Gou}, we propose that the phase shift $\Delta\phi$ these mode will acquire upon propagation from $z=0$ to the focusing plane $z=\infty$ will be given by,
\begin{ceqn}
\begin{equation}
   \Delta\phi=\frac{|m_2|-|m_1|}{m_2-m_1}\frac{\pi}{2},
   \label{Gouy2}
\end{equation} 
\end{ceqn}
where we have used that $\Delta\xi=\arctan(z/z_R)=\pi/2$.

\begin{figure}[h!]
    \centering
    \includegraphics[width=0.98\textwidth]{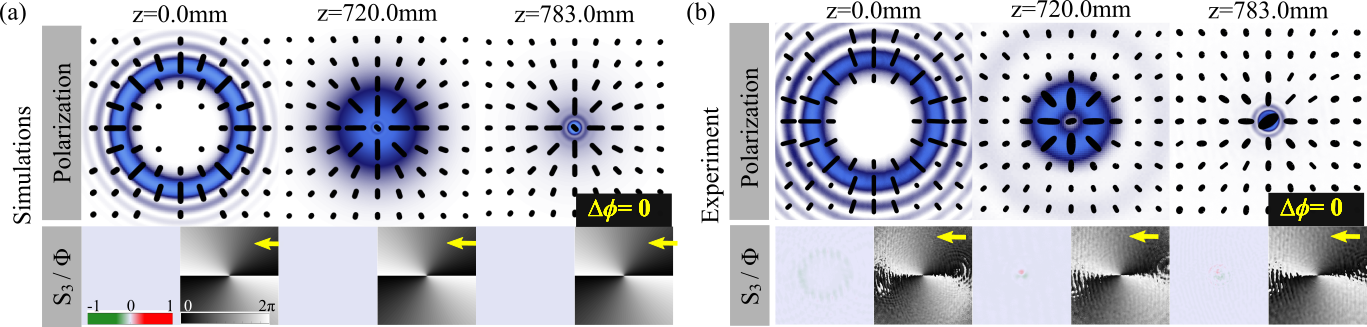}
    \caption{(a) Simulations and (b) experimental propagation dynamics of  $\Psi_{1,-1}$ modes. Top to bottom: intensity overlapped with polarization distribution and $S_3$ parameter (left) and the intermodal phase $\Phi=S_1+iS_2$ (right). Such phase rotate an angle of $\Delta\phi=0$ which is marked as yellow arrow.}
    \label{Result1}
\end{figure}

As a first example we present the cases with equal topological charges $\lvert m_1 \rvert =\lvert m_2 \rvert$, which is shown in Fig. \ref{Result1} for the specific case $\Psi_{1,-1}$. Here, numerical simulations are shown in Fig.\ref{Result1} (a) while experimental results in Fig.\ref{Result1} (b). The first row of each figure shows the polarization distribution overlapped with the intensity distribution. The second row shows the reconstructed Stokes parameter $S_3$ (left panel) and the intermodal phase (right panel). Note that upon propagation the intensity distribution undertakes an abrupt autofocusing effect, as expected. Notice also that the polarization distribution remains invariant, always featuring a radial distribution. In addition, the Stoke parameter $S_3$ also features no change. Finally, given that the absolute value of both topological charges is the same, the Gouy phase is in general $\Delta \phi=0$ and therefore, no phase rotation will be observed. 

\begin{figure}[h!]
    \centering
    \includegraphics[width=0.98\textwidth]{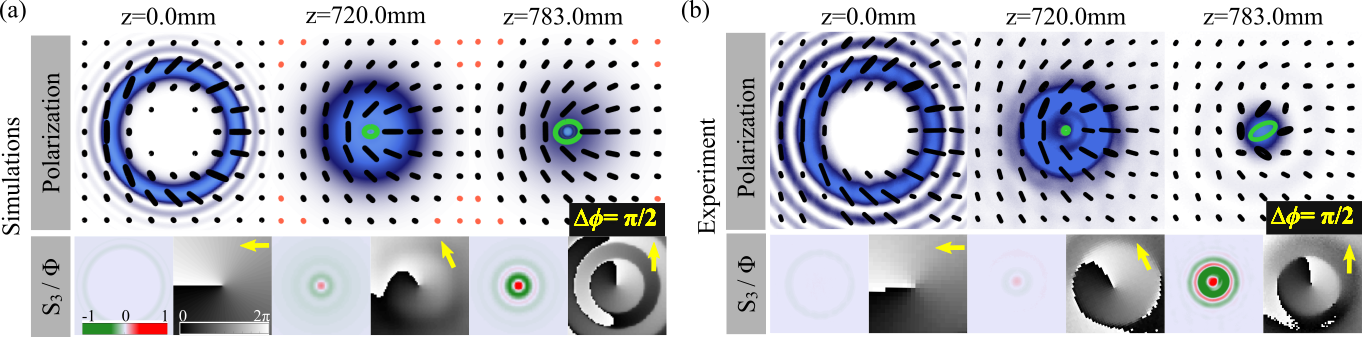}
    \caption{(a) Simulations and (b) experimental propagation dynamics of  $\Psi_{1,0}$ modes. Top to bottom: intensity overlapped with polarization distribution and $S_3$ parameter (left) and the intramodal phase $\Phi=S_1+iS_2$ (right). Such phase rotate an angle of Gouy phase $\Delta\phi={\pi}/{2}$, which is marked as yellow arrow in the inset. Noteworthy, we show the right and left polarization with orange and green ellipses, while the linear polarization black lines.}
    \label{Result2}
\end{figure}

We now present results regarding the cases with different topological charges $|m_1|>|m_2|$, using the specific case $\Psi_{1,0}$ shown in  Fig.\ref{Result2}, numerical simulations in (a) and experimental results in (b), as an example. Again, the reconstructed polarization distribution overlapped with the intensity profiles are shown on the top panels of each figure, whereas the Stokes parameter $S_3$ and intermodal phase $\alpha=\arctan(S_2/S_1)$ on the left- and right-hand side, respectively, of bottom panel. Again, and as expected the intensity profile experiences an abrupt autofocusing propagation behaviour. Further, at $z=0$ this mode features a polarization distribution of a mixed linear polarization states, which upon propagation evolve to elliptical. This can be particularly observed from the Stokes parameter $S_3$, that precisely allows to quantify the amount of right- and left-circular polarization, that in the centre of the beam increases from 0 to 1. This transition can be explained as a transition from OAM to Spin Angular Momentum(SAM). In regards to the intermodal phase, the fact that $|m_1|\neq|m_2|$ causes, for this specific case a clockwise phase rotation $\Delta\phi=\pi/2$ as shown at the  bottom of each figure. Noteworthy, this phase rotation is in accordance with  Eq.\ref{Gouy2}.   

\begin{figure}[h!]
    \centering
    \includegraphics[width=0.98\textwidth]{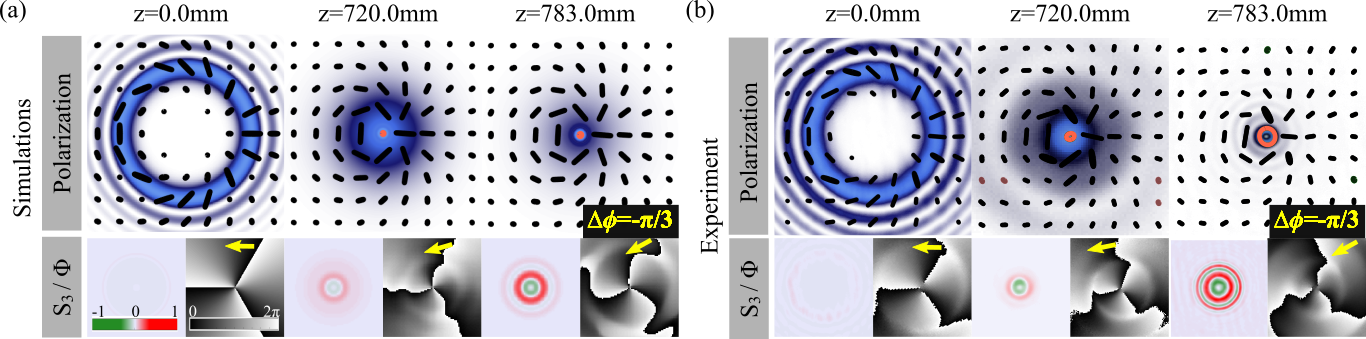}
    \caption{(a) Simulations and (b) experimental propagation dynamics of  $\Psi_{1,-2}$ modes. Top to bottom: intensity overlapped with polarization distribution and $S_3$ parameter (left) and the intermodal phase $\Phi=S_1+iS_2$ (right). Such phase rotate an angle of Gouy phase $\Delta\phi=-{\pi}/{3}$, which is marked as yellow arrow in the inset. Noteworthy, we show the right and left polarization with orange and green ellipses, while the linear polarization black lines. }
    \label{Result3}
\end{figure}

We finally analyse the case of $|m_1|<|m_2|$, whereby and without the loss of generality, the specific case $\Psi_{1,-2}$ is shown in Fig.\ref{Result3}. In a similar way to the previous cases, the transverse polarization distribution overlapped with the intensity profile are shown in Fig.\ref{Result3} (a) and \ref{Result3} (b), numerical simulation and experimental results, respectively. Here again, an abrupt autofocusing behaviour in the intensity profile can be observed as the beam propagates. Additionally, their polarization distribution in the transverse plane also experience a transition from linear to elliptical. This behaviour can be quantified through the Stokes parameter $S_3$ on the bottom-left inset, where the intensity of each circular polarization component (left and right) increases as the beam propagates, in a similar way to the previous case. Finally, here also a phase rotation is observed originated from the inequality of $|m_1|$ and $|m_2|$, which in contrast to the previous case, it happens in an anticlockwise direction, as shown in the bottom-right inset of the same figure. For this specific example, a total phase rotation $\Delta\phi=-{\pi}/3$ can be observed, in accordance to Eq.\ref{Gouy2}.
\section*{Discussion}

In this manuscript we introduced, theoretically and experimentally the Circular Airy Gaussian vortex vector modes, generated as a weighted superposition of Circular Airy Gaussian vortex beams. Such vector modes not only inherit the self-accelerating and autofocusing behaviour from their scalar regime, but also present a propagation dynamics of their transverse polarization distribution and intermodal phase rotation. To be more specific, their transverse polarization distribution evolves from linear to  elliptical along the auto-focusing process, while the intermodal phase original angle allows to be under controlled following an invriant, clockwise and anti-clockwise rotated trajectories. A possible explanation for this lies in the fact of a transfer from OAM to SAM, which can be directly observed via Stokes parameters. To corroborate this, we conduct a details analysis of the propagation dynamics of our generated CAGVV beams at various propagation distances and for three cases, namely $|m_1|=|m_2|$, $|m_1|>|m_2|$ and $|m_1|<|m_2|$. From this, we first reconstructed their transverse polarization distribution via Stokes polarimetry, where a polarization distribution of a mixed linear polarization states evolve to elliptical upon propagation. We further analysis the propagation dynamics of the intermodal phase that CAGVV modes experience upon propagation. From this we observed a propagation-invariant intermodal phase for the case $|m_1|\neq|m_2|$, a clockwise and anticlockwise phase rotation the cases $|m_1|>|m_2|$ and $|m_1|<|m_2|$, respectively. This funding allows us to provide a mathematical expression that related the rotation with both topological charges, which also fits well with previous studies applied to the scalar beams. We further quantified the observed experimental rotation through numerical simulations and demonstrated these results fit very well with the mathematical expression, evincing that indeed, the OAM not only directly affect the intermodal phase rotation, but also can be transferred to SAM upon propagation, see the intensity dynamics undergoes from zero to maximum in the transverse distribution of the Stoke parameter $S_3$. It is also worth mentioning that the dynamics of CAGVV can be adjust through the initial launch angle parameter $\nu$, which allows a wide variety of possibilities in the manipulation of propagation trajectories, as well as their polarization-variant structure.  Finally, these novel vectorial states of light, are of paramount importance, which will pave the way towards the development of novel applications. For example, the CAGVV beams allows to offer a flexible tool for practical applications in a wide variety of fields, such as optical manipulation, optical tweezers, amongst others.
\section*{Methods}
\begin{figure}[h!]
    \centering
    \includegraphics[width=0.78\textwidth]{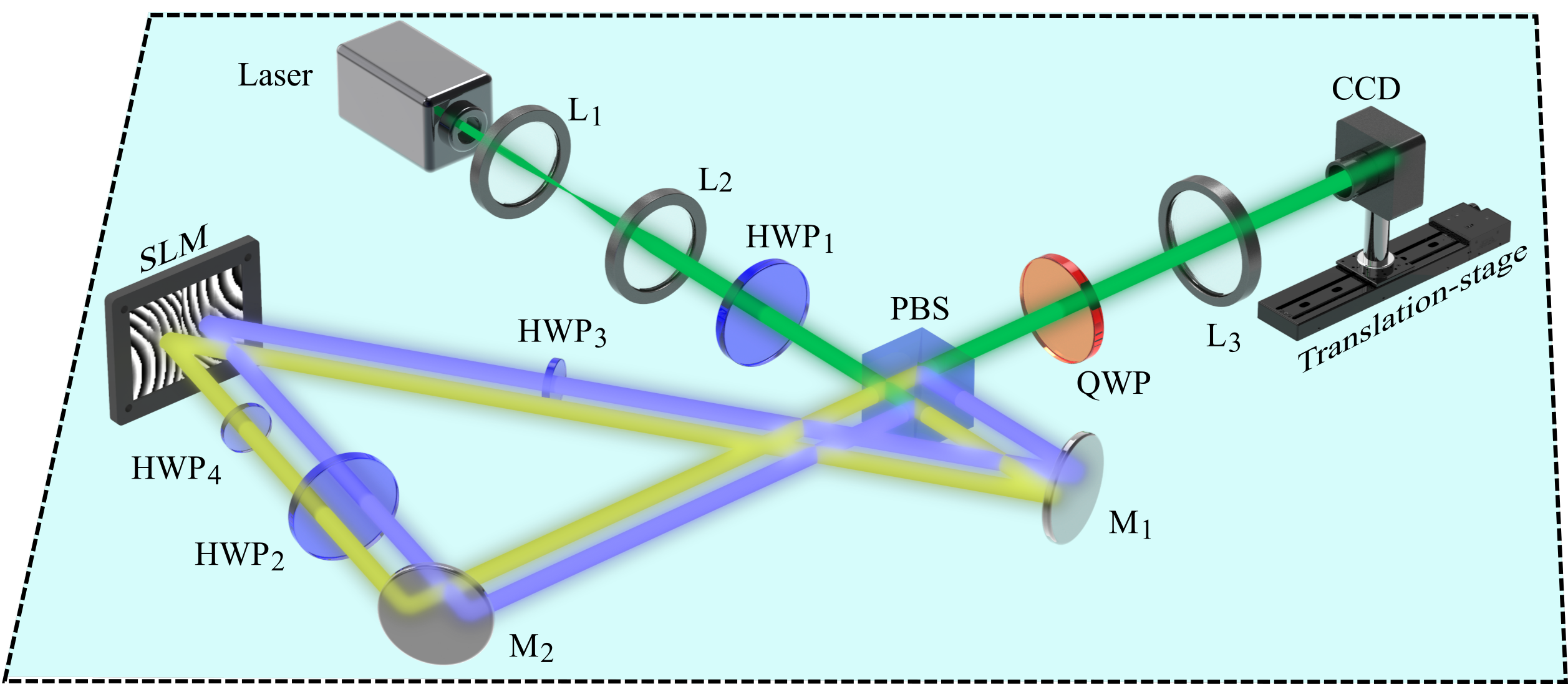}
    \caption{Schematic representation of the optical setup implemented to generate CAVV mode. L$_1$-L$_3$: lenses, HWP$_1$-HWP$_4$: Half-Wave plate, QWP: Quarter-Wave Plate, PBS: Polarizing Beam Splitter, M$_1$-M$_2$: Mirror, SLM: Spatial Light Modulator, CCD: Charged Device Camera.}
    \label{setup} 
\end{figure}
\begin{figure}[tb]
    \centering
    \includegraphics[width=0.89\textwidth]{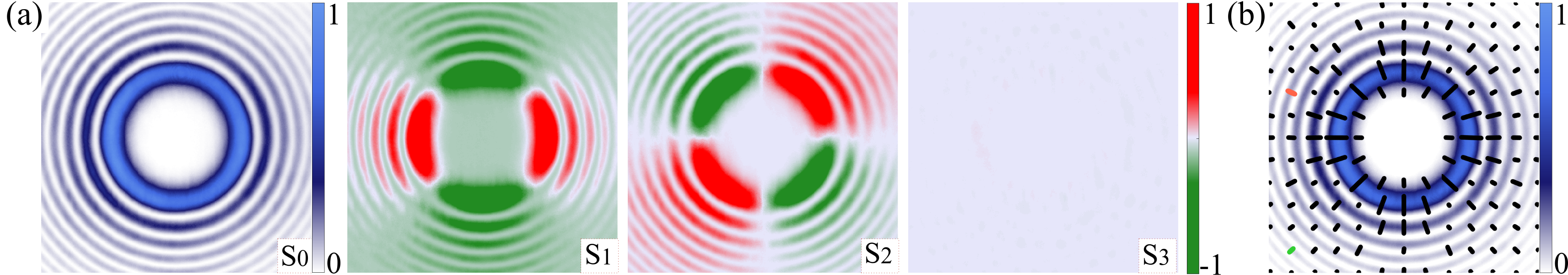}
    \caption{Experimental Stokes parameters $S_{0}$, $S_{1}$, $S_{2}$ and $S_{3}$ of the vector mode $\Psi_{1,-1}$ are represented in (a), respectively, from which the polarization distribution can be reconstructed, as shown in (b), where the intensity profile is also shown.}
    \label{Stokesrecons} 
\end{figure}

To generate CAGVV beams we implemented the experimental setup schematically depicted in Fig.\ref{setup}. Here, a linear horizontally polarized laser beam ($\lambda=532$nm) is expanded by lenses L$_1$ and L$_2$, and subsequently transformed into a diagonally polarized beam with the use of a half-wave plate (HWP$_1$) at $22.5^\circ$. Afterwards, a polarizing beam splitter (PBS) separates the beam into its horizontal (yellow color) and vertical (purple color) polarization components. With the help of mirrors (M$_1$ and M$_2$), both beams are directed to a spatial light modulator (SLM), which in conjunction with the PBS for a highly stable interferometer. The screen of SLM is digitally split into two sections (left and right), each of which is addressed with one independent hologram, which encodes the Fourier transformation of the constituting wave fields defined by Eq.\ref{Eq:CAV}. In combination with a lens L$_3$ ($f=300$mm), two independent optical fields are generated in the focus plane. Due to the fact that SLMs only modulate horizontal polarization, HWP$_{2,3,4}$ are used to properly modulate the polarization states of both beams, before and after hitting the SLM. Notice that each arm travels the same optical path, no extra phase difference is induced in the whole round trip. Finally, the two target optical fields exit the interferometer from the opposite side of the PBS, travelling along a common-path, generating in this way the desired vector beams in the linear polarization basis. To transform them from the linear to the circular polarization basis, a quarter-wave plate (QWP) at 45$^\circ$ is inserted along the path of the beam.  Finally, a charged-coupled device (CCD: FL3-U3-120S3C-C with a resolution of 1.55$\mu$m) placed in the focal plane of L$_3$, is mounted on a rail to record the intensity of the generated beams as a function of propagation. From such intensity measurements the polarization distribution of CAGVV modes are reconstructed, directly from the Stokes Parameters $S_0, S_1, S_2,$ and $S_3$, through the relations \cite{Rosales2022},
\begin{ceqn}
\begin{equation}\label{Eq.SimplyStokes}
\begin{split}
\centering
 S_{0}=I_{H}+I_{V},\hspace{8mm} S_{1}=2I_{H}-S_{0},\hspace{8mm}
 S_{2}=2I_{D}-S_{0},\hspace{8mm} S_{3}=2I_{R}-S_{0}.
\end{split}
\end{equation}
\end{ceqn}
Here, the required intensity projections, which represent the horizontal, vertical, diagonal and right-circular polarization components are represented by $I_{H}$, $I_{V}$, $I_{D}$ and $I_{R}$, respectively. Such intensities can be experimentally measured through a series of phase retarders inserted in front of the CCD camera. More specifically, we measure the $I_{H}$, $I_{V}$ and $I_{D}$ by passing the generated beam through a polarizer oriented at $0^\circ$, $90^\circ$ and $45^\circ$, respectively, while $I_{R}$ is measured by inserting a QWP at $45^\circ$ with the polarizer at $90^\circ$ (see [\cite{Rosales2022}] for more details). As way of example, we show the experimental Stokes parameters in Fig.\ref{Stokesrecons}(a), where the reconstructed polarization distribution overlapped with the intensity profile $S_{0}$ for the specific case $\Psi_{1,-1}$ at the plane $z=0$ is represented in Fig. \ref{Stokesrecons}(b). It is also worth noting that the same amount of right- and left- circular polarized components in this case results in the zero intensity distribution of the Stokes parameter $S_{3}$ (see right bottom panel in Fig.\ref{Stokesrecons}(a)), that is also the reason that the final reconstructed transverse polarization distribution features with radial linear polarized states (see Fig.\ref{Stokesrecons}(b)). 
\bibliography{References}
\section*{Acknowledgements}

This work is supported by Science Foundation of Zhejiang Sci-Tech University (ZSTU) under Grant No. 22062025-Y and the National Natural Science Foundation of China (61975047).

\section*{Author contributions statement}
 B.Z performed the experiments. X.B.H, B.Z and C.R.G. analysed and interpreted the results. X.B.H and C.R.G. wrote the manuscript. X.B.H, R.P.C and C.R.G. supervised the project.

\section*{Additional information}
Competing Interests: The authors declare no competing interests.

\end{document}